\documentclass[runningheads]{llncs}
\usepackage{graphicx}
\usepackage{diagbox}
\usepackage[ruled,vlined]{algorithm2e}
\usepackage{multirow}
\usepackage{xcolor}
\usepackage{amsmath}

\begin{document}

\title{ProphetNet-Ads: A Looking Ahead Strategy for Generative Retrieval Models in Sponsored Search Engine}

\titlerunning{ProphetNet-Ads}

\author{
Weizhen Qi\inst{1} \thanks{Work is done during internship at Microsoft Research Asia.} \and
Yeyun Gong\inst{2} \and
Yu Yan\inst{3} \and
Jian Jiao\inst{3} \and
Bo Shao\inst{4} \and
Ruofei Zhang\inst{3} \and
Houqiang Li\inst{1} \and
Nan Duan\inst{2} \and
Ming Zhou\inst{2}
}

\authorrunning{W. Qi et al.}

\institute{
University of Science and Technology of China, Hefei, China \\ \email{\{weizhen, lihq\}@mail.ustc.edu.cn} \and
Microsoft Research Asia, Beijing, China \\ \email{\{yegong, nanduan, mingzhou\}@microsoft.com } \and
Microsoft, Redmond, USA \\ \email{\{yyua, jian.jiao, bzhang\}@microsoft.com} \\  \and
Sun Yat-sen University, Guangzhou, China \\ \email{shaobo2@mail2.sysu.edu.cn}
}

\maketitle              % typeset the header of the contribution
\begin{abstract}
In a sponsored search engine, generative retrieval models are recently proposed to mine relevant advertisement keywords for users' input queries. Generative retrieval models generate outputs token by token on a path of the target library prefix tree (Trie), which guarantees all of the generated outputs are legal and covered by the target library. In actual use, we found several typical problems caused by Trie-constrained searching length. In this paper, we analyze these problems and propose a looking ahead strategy for generative retrieval models named ProphetNet-Ads. ProphetNet-Ads improves the retrieval ability by directly optimizing the Trie-constrained searching space. We build a dataset from a real-word sponsored search engine and carry out experiments to analyze different generative retrieval models. Compared with Trie-based LSTM generative retrieval model proposed recently, our single model result and integrated result improve the recall by 15.58\% and 18.8\% respectively with beam size 5. Case studies further demonstrate how these problems are alleviated by ProphetNet-Ads clearly.

\keywords{Sponsored Search Engine \and Generative Retrieval Model \and  Keywords Extension \and Information Retrieval \and Natural Language Generation  }
\end{abstract}
\section{Introduction}
\begin{figure}[h]
    \centering
	\includegraphics[width = 4.5in]{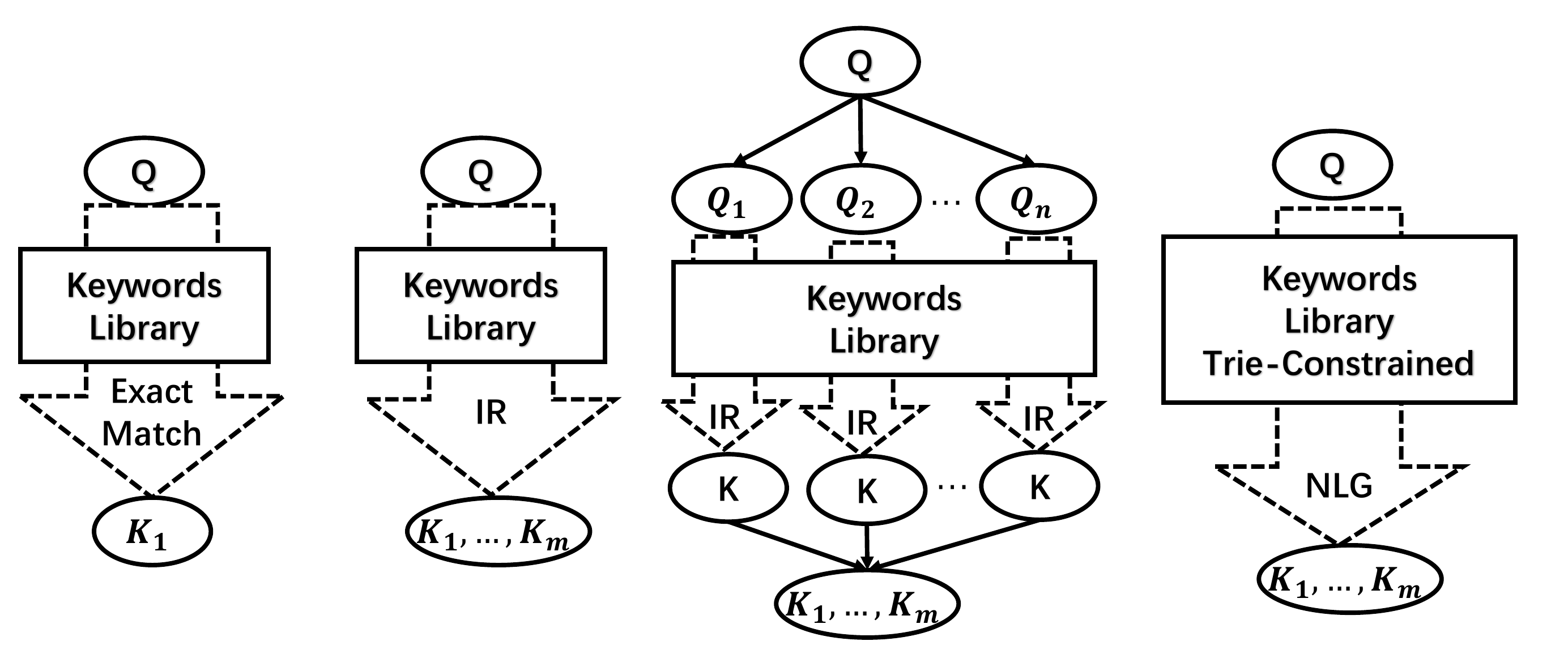}
	\caption{The models used for keywords extension task. Firstly, triggered ads keywords are the exact match of users' query. Secondly, information retrieval techniques are used to mine similar ads keywords. Thirdly, users' queries are re-written into intermediate queries to do IR, which alleviate the semantic gap between queries and ads keywords. Recently, generative retrieval models are used for keywords extension task, which ensures all the generated keywords all covered by the target library because of the constrained searching space.  }\label{intro.task}
\end{figure}

In a sponsored search engine, search queries from the user are expanded to appropriate advertisements (Ads) keywords. Advertisers bid on triggered keywords to display their ads and pay by click. The primary income for a sponsored search engine is to provide ads that users potentially need. Therefore the applications of keywords extension from queries to relevant keywords in the ads library are deeply concerned. At the beginning, search engines trigger ads when the queries are identical with an ads keyword. Then, methods like Information retrieval (IR) with quality filter~\cite{hillard2010improving} are commonly used to recall more relevant keywords. However, traditional IR techniques are unable to fill the semantic gap between queries and ads keywords. Thus sponsored search engines pay much attention on how to excavate more semantic-related keywords. A solution is to re-write the initial user queries to a range of intermediate queries and then combine all the outcomes retrieved from them, such as \cite{jones2006generating} from Yahoo, \cite{riezler2010query} from Google, and \cite{gao2012learning} from Microsoft. Re-writing strategies are widely used because directly extending queries to keywords will lead to the low-efficiency problem: very few extensions are included in the keywords library. Recently ~\cite{lian2019end} used Trie-based LSTM model to address this problem by constraining the generation searching space. Trie means a prefix tree. Trie-based NLG models generate tokens on paths of a Trie to make sure outputs are covered by the keywords library.The models used for keywords extension task in different stages are shown in Figure\ref{intro.task}.

\begin{figure}[h]
    \centering
	\includegraphics[width = 4.5 in]{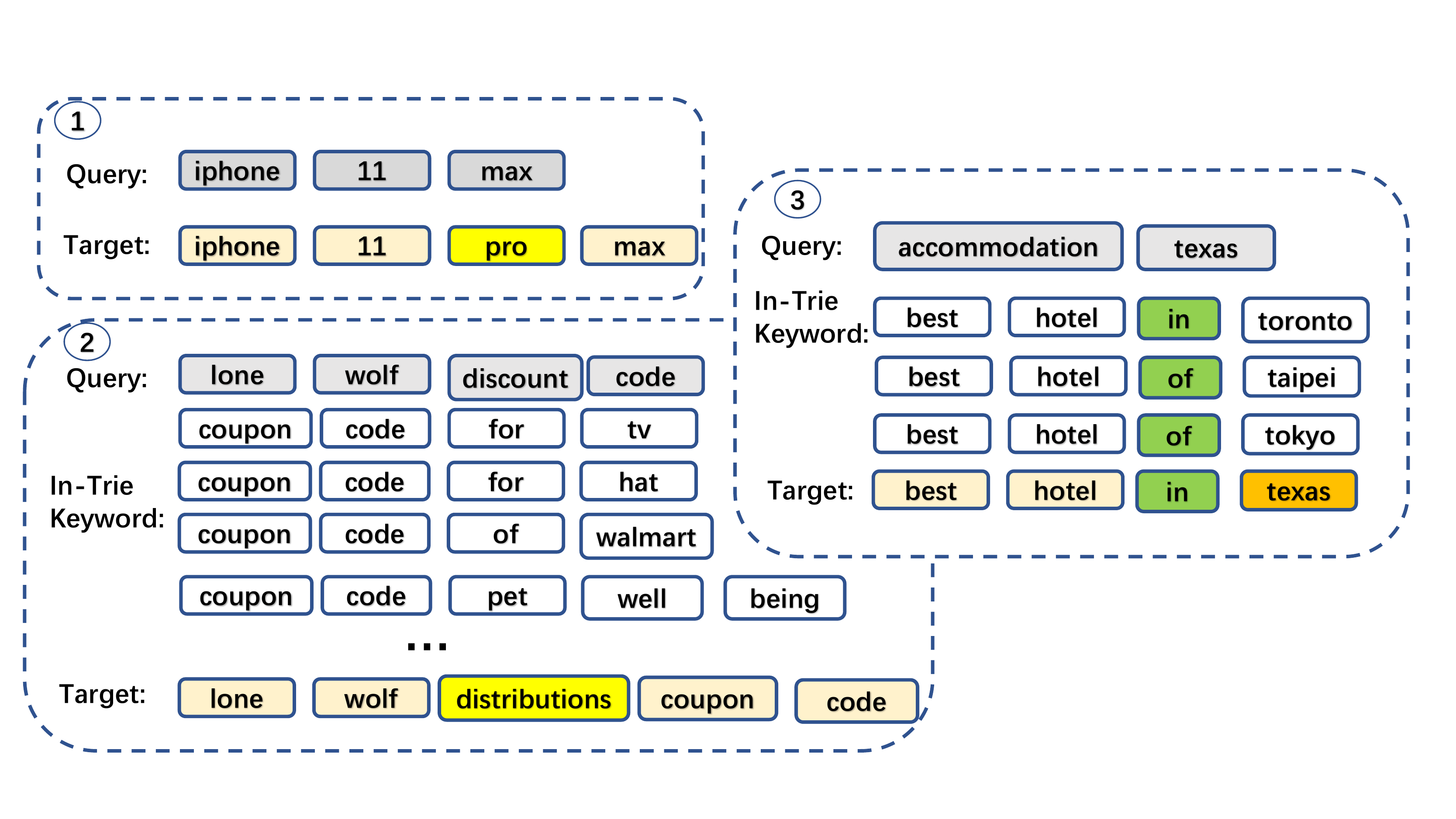}
	\caption{In  example 1, "pro" is a noise word with low generation score if NLG model is not trained on this data. In example 2, generative retrieval models will be easily trapped into the common prefix "coupon code xxx". In example 3, both "in" and "of" have high generation scores but "of" has no desired suffix "texas'.}\label{intro.prob}
\end{figure}

However, simply adding a Trie constraint to a natural language generation (NLG) model is not enough, and we found several common problems in daily use. The biggest problem of Trie-based generative retrieval model is that it cannot utilize global information. We list three examples in Figure \ref{intro.prob}.  The first problem is that noise tokens will have a very low generation score thus lead to a wrong searching path. A second common problem is called "common prefix has no target object in the future tokens", which implies that the entire beam search space is filled with common prefixes. Although these prefixes may compose good keywords, sometimes expected suffixes are not in the Trie to compose desired keywords.  We cannot simply throw these sequences from beam search unfinished queue as these prefixes are really "common" and take a portion of good results. The last problem is that models are hard to decide which one is better when several tokens have similar high generation scores. For keywords extension, models have no idea which suffix will lead to desired keywords in a Trie.

Inspired by ProphetNet\cite{yan2020prophetnet}, which is able to predict next several tokens simultaneously, we propose ProphetNet-Ads to alleviate above problems with the future information. ProphetNet is proposed as a new pre-training architecture to predict next n-grams. % to alleviate local dependency overfitting and enhance the future information. 
%However, although next n-grams can be predicted, only the next first token of ProphetNet is used in inference procedure like traditional NLG models. 
ProphetNet-Ads employs the future tokens' scores to look ahead several steps in the Trie, which directly optimizes the searching space. With Trie-based beam search, the next token to generate is constrained to possible suffixes of the decoded hypothesis according to the Trie. ProphetNet-Ads is proposed for better selection of the suffixes. ProphetNet-Ads modifies the predicting tokens' scores as a weighted sum of its generation score and future information scores to optimize the searching space. We rank the decoding hypothesis with the modified scores, but store the unchanged sentence scores, which optimizes searching space and meantime keeps the scores consistent to original NLG model. The experimental results show that our proposed strategies recall more relevant keywords with an obvious improvement. Case studies further demonstrate how ProphetNet-Ads alleviates these typical problems.

\section{Background}
\noindent\textbf{ProphetNet} 

\noindent ProphetNet\cite{yan2020prophetnet} is recently proposed as a new pretraining NLG architecture. To alleviate strong local correlations such as bi-gram combination and enhance the hidden states to contain more global information, next n-grams are trained to predict. ProphetNet employs n-stream self-attention to support next n-grams from any starting positions in a given output are trained to predict simultaneously. Although next n-grams are explicitly used in the training procedure, only the next first token is predicted in the inference procedure like traditional NLG models. These future tokens' scores can be used to point out whether the next first token has desired information in a Trie. 

\noindent\textbf{Trie-based NLG}

\noindent A Trie is a prefix tree, and a path from the starting token to an internal node denotes a prefix of a sequence, a path from the starting token to a leaf node denotes a complete sequence. Suppose the already decoded token sequence is a prefix of a legal keyword sequence, then it must be a route in Trie, and we generate next tokens from the suffix nodes of this route. In this manner, all of the generated outputs are in-library. Trie-based inference have been successfully used in NLG tasks in recent years~\cite{kannan2016smart,ye2018photoreply,laddha2019understanding,lian2019end}.~\cite{kannan2016smart} firstly used Trie to constrain the model output candidates for email replying task. It can also been seen as picking responses from already given sentences in a Trie for any given email. 

\noindent\textbf{Keywords Extension for Sponsored Search Engine}

 \noindent Sponsored search engine service providers are deeply concerned with the task of extending users' input queries into ads keywords. Researches are carried out to fill the semantic gap between queries and ads keywords. One solution is to re-write the initial user queries to intermediate queries to retrieve keywords, such as \cite{gao2012learning,jones2006generating,riezler2010query}. With the improvement of NLG techniques, ~\cite{he2016learning} used LSTM to train the re-writing model, utilizing the deep learning network for better semantic modeling ability. ~\cite{lee2018rare} from Microsoft directly trained a NLG model to generate candidate ads keywords. Even though the NLG model's outputs are highly qualified, however, they have a high likelihood to be out of the target set. Recently ~\cite{lian2019end} used Trie-based NLG model to overcome the low-efficiency barrier by restricting the search space, and this methodology brought a considerable enhancement for their system with an additional 10\% revenue each year. 

\section{ProphetNet-Ads}

\begin{figure}[h]
    \centering
	\includegraphics[width = 4.5 in]{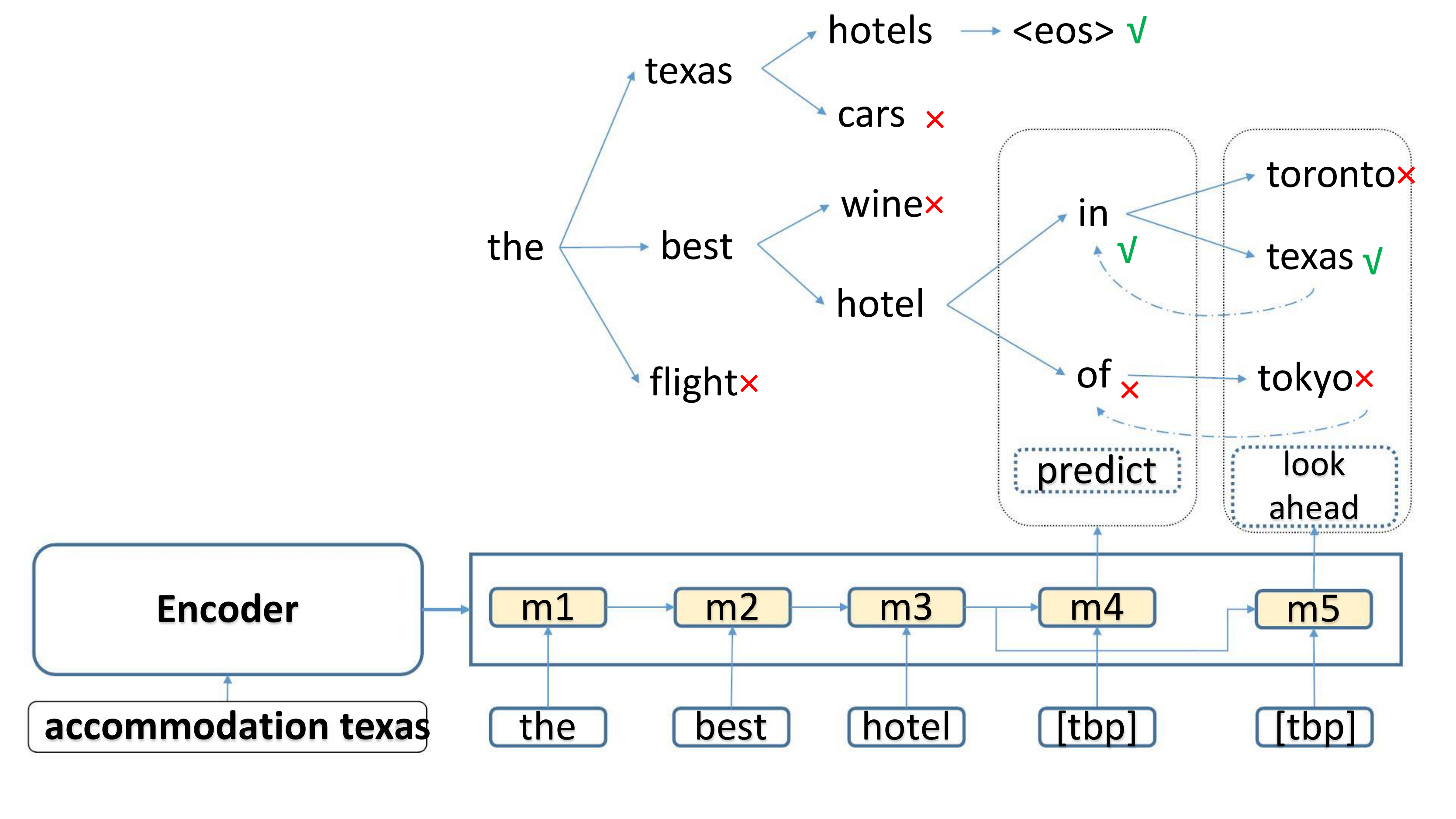}
	\caption{An example of Bi-gram ProphetNet-Ads. When generating next token for "the best hotel", "in" and "of" are its suffix tokens according to the Trie. Though both of them are good suffixes to generate, "of" has no future tokens with high score, while future tokens of "in" cover desired token "texas". Thus "in" is generated.
	}\label{trie.decode}
\end{figure}

Based on ProphetNet which is able to predict more future tokens, we propose an explicit looking ahead strategy named ProphetNet-Ads as a possible solution for problems discussed in the introduction. ProphetNet-Ads modifies the scores of the next first predicting tokens by looking ahead future tokens' scores and directly optimizes the searching space.  Figure\ref{trie.decode} shows an illustration of ProphetNet-Ads generation procedure.

ProphetNet-Ads modifies the in-Trie suffix tokens' scores with the information of its future tokens when beam searching on a Trie. We look ahead $\ell$ steps, where $\ell$ is usually $n-1$ for a ProphetNet $n$-gram generation model, since we can generate $n$ tokens simultaneously, the next first predicting token, and $n-1$ future tokens to look ahead for this suffix. A residual weight $\lambda$ is set to control the weight of next token's generation score and its looking ahead score.

As shown in Figure \ref{trie.decode}, a Bi-gram ProphetNet is able to generate next two tokens' generation scores at each time step, and we can call them  $g_1, g_2$. We refer the previous decoded sequence as $seq$, and next first suffixes of $seq$ as $s1$. For each node $\rho_1$ in $s1$, one step further suffixes of $\rho_1$ are noted as $s2$. The generation score of next first token $\rho_1$ is modified as:

\begin{equation}
g_1[\rho_1] = \lambda \times g_1[\rho_1] +(1 - \lambda) \times max(g_2[s_2])
\end{equation}
    
For example, the step scores for the suffixes we are predicting from Figure\ref{trie.decode} are modified as:

\begin{equation}
\begin{aligned}
g_1["in"] &= \lambda \times g_1["in"] + (1 - \lambda) \times max(g_2["toronto"], g_2["texas"]) \\
g_1["of"] &= \lambda \times g_1["of"] +(1 - \lambda) \times g_2["tokyo"] 
\end{aligned}
\end{equation}

\begin{algorithm}
\SetAlgoLined
\SetKwInOut{Input}{input}\SetKwInOut{Output}{output}
\Input{Beam Size $b$, n-gram ProphetNet $\mathcal{P}$, Trie $T$, Residual weight $\lambda$, Input query $X$, max output token length $l$}
\Output{Keywords extensions $\pi$}
 alive buffer: $\mathcal{H} \leftarrow \emptyset$ ; 
 finished buffer: $\pi \leftarrow \emptyset$ \ \tcp*{ with [hypothesis, scores]}
 put [bos, score\_bos] in $\mathcal{H}$ \tcp*{Initialize the alive buffer}
 \While{$best\_alive\_score \geq worst\_alive\_score$ and $ decoded\_length < l$}{
  $\mathcal{O}_{sen} \leftarrow \emptyset$  \tcp*{Original sentence scores to be stored in $\mathcal{H}$}
  $\mathcal{M}_{sen} \leftarrow \emptyset$  \tcp*{Modified sentence scores to be ranked temporarily}
  \For{$seq$ in $\mathcal{H}$}{
    [$g_1$,$g_2$,...,$g_n$] $\leftarrow \mathcal{P}(seq,X)$ \tcp*{Next future n tokens' scores}
    $s_1,m_1 \leftarrow T(seq)$ \tcp*{$s1$: suffix tokens, $m1$: mask vector}
    $\mathcal{O}_{token} = \mathcal{M}_{token} = g_1 + m_1$ \tcp*{Mask the tokens out of Trie}
    \For{$\rho_1$ in $s_1$ \tcp*{Start looking ahead}  }{ 
    $s_2,m_2 \leftarrow T(seq + \rho_1)$\;
    \For{$\rho_2$ in $s2$  \tcp*{Could be replaced with recursive function}  }{
        $s_3,m_3 \leftarrow T(seq + \rho_1 + \rho_2)$ \;
        \For{$\rho_{...}$ in $s_{...}$      }{
            ...\;
            \For{$\rho_{n-1}$ in $s_{n-1}$}{
            \tcp{Modify scores from the farthest nodes}
                $s_n, m_n \leftarrow T(seq + \rho_1 + \rho_2+...+\rho_{n-1})$\;
                $g_{n-1}[\rho_{n-1}]= \lambda  \times g_{n-1}[\rho_{n-1}] + (1-\lambda)\times max(g_n+m_n)$ \;
            }
            ...\;
        }
        $g_{2}[\rho_2]= \lambda  \times g_{2}[\rho_2] + (1-\lambda)\times max(g_3+m_3)$\;
    }
    \tcp{Modify scores until the next first token}
    $\mathcal{M}_{token}[\rho_1] =  \lambda \times \mathcal{O}_{token}[\rho_1]  + (1-\lambda)\times(max(g2+m2))$\;
    }
    \tcp{Calculate new sentence scores with previous decoded score and next first tokens' step score}
    $\mathcal{O} \leftarrow func(seq.score, \mathcal{O}_{token})$ put $\mathcal{O}$ into ${O}_{sen}$  \tcp*{Original scores}
    $\mathcal{M} \leftarrow func(seq.score, \mathcal{M}_{token})$ put $\mathcal{M}$ into ${M}_{sen}$  \tcp{Modified scores}
  }
  \tcp{Rank with modified scores but store their original scores}
  $new\_seqs, id \leftarrow top\_b\_of(\mathcal{M}_{sen})$ \;
  $new\_finished\_seqs, id\_f \leftarrow top\_b\_of(\pi.socres, \mathcal{M}_{sen}.eos)$ \;
  $\mathcal{H} \leftarrow new\_seqs, \mathcal{O}_{sen}[id] $ \;
  $\pi \leftarrow new\_finished\_seqs, \mathcal{O}_{sen}[id\_f] $;
 }
 return  $\pi$\;
 \caption{N-gram ProphetNet-Ads Trie-based Searching}
 \label{alg.beam}
\end{algorithm}

Similarly, a n-gram generation model could output the probability distributions of next $n$ tokens as $g_1,g_2,...g_n$. We use a recursive function to modify their scores from the furthest to the nearest next first tokens' scores. Scores of $g_{n-1}$ are modified with their highest children nodes' scores in $g_n$, and then be used to modify $g_{n-2}$, until next first tokens' scores $g_1$ are modified. Then, the best token in $g_1$ is chosen. Considering a high-confidence suffix before explicit looking ahead strategy, if it has no good tokens steps further, a low future score will be passed backward. On the opposite if there are any noise tokens in suffix but with expected tokens in the future, further high-confidence scores will also be passed across the noise to give a bonus for the token we are predicting.

However, if we directly use the modified generation tokens' score $g_1$ to calculate decoded sequence scores in beam search, results are inconsistent with the generation model as it modifies the output sequences scores, which could bring error accumulation. Thus, we only use the modified scores to rank and pick the best sequences, but store their original scores. ProphetNet-Ads not only optimizes the searching space but also keeps the scores consistent to the generation model. The algorithm of ProphetNet-Ads is described in Algorithm\ref{alg.beam}.

\section{Experiment}
In this section, we introduce the dataset and implementations of models to validate ProphetNet-Ads. Since ProphetNet only releases its Bi-gram uncased pretrainied checkpoint\footnote{https://github.com/microsoft/ProphetNet} 
for now, for fair comparison, in this paper the Uni-gram to Tri-gram ProphetNet or ProphetNet-Ads are finetuned in ProphetNet architecture but without pretraining.   

\subsection{Dataset}

 The keywords extension dataset is collected from Bing search engine keywords extension library, formed as "query, triggered ads keyword" pairs. They are collected from advertisers, human labelling, searching log history, high quality extensions from old algorithms, etc. 260 million keywords are used to build a Trie as the searching space. After a quality model and Trie-filtering, we randomly select one million high-qualified training data and ten thousand testing data. The average length for target keywords after WordPiece tokenization is 6.69 and the average length for training data query input is 4.47. Each query from the testing data has at least one associated ads keyword, but we are unsure of how many other related keywords it has in the Trie. In actual use for a sponsored search engine, a number of relevant keywords are generated for a given query for further filtering and subsequent processing. More relevant keywords are recalled is concerned. Under this setting, we use recall rate to compare different models. MAP(mean average precision) is also included for comparison in the main results Table~\ref{tb:result.overallresult}.

\subsection{Model Settings}
We implement both traditional IR algorithm BM25 and a list of generative retrieval models as our baseline. Okapi BM25 \cite{robertson1995okapi} is a traditional IR strategy, with the word tokenization of nltk\cite{loper2002nltk} and parameters as $k_1=1.2,b=0.75,\epsilon=0.25$. Second type baseline is Trie based LSTM models as proposed by \cite{lian2019end}. A 4-layer encoder, 4-layer decoder uni-directional LSTM+Trie model is implemented according to the complex model for offline use of \cite{lian2019end}. Improvements are added based on it. We change the uni-directional LSTM encoder to bi-directional LSTM encoder to validate the effects of encoding bi-directional information. Copy mechanism\cite{gu2016incorporating,see2017get} gives a bonus to generation scores of those words appear in the input sequence. Output keywords often have some overlap with the input queries, and copy mechanism allows model to directly pick some tokens from the input to compose the answer, which improves the generation ability for overlapped tokens. We train ProphetNet\_large~\cite{yan2020prophetnet} models with copy mechanism as the third baselines. ProphetNet-Ads shares the same checkpoint as ProphetNet baselines, with additional proposed optimization by looking ahead.

All generative retrieval models use a same 30,000 words vocabulary with WordPiece~\cite{wu2016google} tokenization and share the same Trie. The LSTM based models are implemented according to~\cite{see2017get}, and trained for 10 epochs. ProphetNet and ProphetNet-Ads are implemented according to~\cite{yan2020prophetnet}, trained with learning rate 3e-4, 5 epochs. Other hyper-parameters are same to the referenced models. Training batch sizes are all set to 36, with a maximum input token length of 20 and a maximum output length of 20. 

\subsection{Results Analyze}

\begin{table*}[ht]
\small\addtolength{\tabcolsep}{-0.01pt}
\begin{tabular}{|l|c|c|c|c|c|c|c|c|c|c|c|c|}
\hline
Model            & R@5   & R@10  & R@15  & R@20  & MAP@5  & MAP@10 & MAP@15 & MAP@20 \\  \hline
BM25            & 27.86 & 33.40 & 37.30 & 39.13  & 0.2051 & 0.2125 & 0.2156 & 0.2166 \\
LSTM             & 62.47 & 71.81 & 75.63 & 77.76  & 0.5716 & 0.6267 & 0.6442 & 0.6534 \\
Bi-LSTM         &  63.28  & 72.28 & 76.21 & 78.13  & 0.5770 & 0.6292 & 0.6479 & 0.6563 \\
Bi-LSTM+Copy         & 67.37 & 76.12 & 79.40  & 83.37 & 0.6114 & 0.6616 & 0.6755 & 0.6811 \\
Uni-gram ProphetNet        & 75.00 & 82.50 & 84.90 & 86.50 & 0.6929 & 0.7362 & 0.7461 & 0.7526 \\ 
Tri-gram ProphetNet  & 75.48 & 83.08 & 85.45 & 86.68  & 0.6974 & 0.7426 & 0.7518 & 0.7565 \\ \hline
ProphetNet-Ads  & 78.05 & 84.28 & 86.24 & 87.54  & 0.7133 & 0.7472 & 0.7542 & 0.7580 \\ 
Merged Tri+Tri-Ads   & 81.34 & 86.83 & 88.45 & 89.39 & /      & /      & /      & /      \\
Merged Above   & 86.56 & 90.11 & 91.34 & 92.15 &  /      & /      & /      & /     \\
\hline
\end{tabular}
\caption{Comparison with traditional IR algorithm BM25 and generative retrieval models. Results include recently proposed Trie-based LSTM model and its enhanced variants, ProphetNet generative retrieval model and ProphetNet-Ads. ProphetNet-Ads uses same checkpoint as Tri-gram ProphetNet, with looking ahead optimization. Merged Tri+Tri-Ads means the results merged with Tri-gram ProphetNet and ProphetNet-Ads. R@x for generation model means recall of generation procedure with beam size x, for BM25 means recall of top x of the IR results.}\label{tb:result.overallresult}
\end{table*}

 We analyze different keywords extension models according to the results in Table~\ref{tb:result.overallresult}. Firstly, we can easily draw the idea that traditional IR algorithm like BM25 is not suitable for keywords extension task, since it cannot fill the semantic gap. Compared with LSTM with the beam size 5, replacing encoder with bi-directional LSTM could improve the recall by 0.81\% and adding copy mechanism could improve the recall by 4.09\% further.  Copy mechanism enhances the results obviously, because the keywords are likely to cover some same words as the input query, copy mechanism enables model to directly fetch some words or word pieces from the input, which is a strong assistant to our model. Compared to the LSTM variants, Uni-gram ProphetNet which is similar to Transformer, improves recall by 7.63\%. This is mainly because the stacked Transformer architecture are deeper and keywords extension task has a big training corpus, with a large amount of features and information for the generation model to capture and learn. Tri-gram ProphetNet improves the recall by 0.48\%, which shows that trained to predict more future tokens helps NLG ability even the future tokens are not explicitly used. ProphetNet-Ads uses the same trained model as Tri-gram ProphetNet, and improves the recall by 2.57\% further. This shows that optimizing searching space in the inference procedure could help a generative retrieval model a lot, and our proposed looking ahead strategy can optimize it effectively by incorporating future information. Merged result is more concerned by sponsored search engine for offline use. From the merged results we observe that, with the same one million training data, integrating different searching space optimization models can generate more satisfactory results.

With the comparison between our models and the baseline models, we see that our proposed looking ahead strategies improve the results obviously. It shows that simply using Trie to constrain the searching space is not enough, and our looking ahead strategies can optimize the searching space and help the keywords extension task effectively. 

\subsection{Ablation Analyze}\label{chapt.parameters}
In this part, we will analyze the choice of how many tokens to predict as the $n$ for n-gram ProphetNet-Ads and the choice of residual weight $\lambda$.

\begin{figure}[h]
    \centering
	\includegraphics[width = 4.5 in]{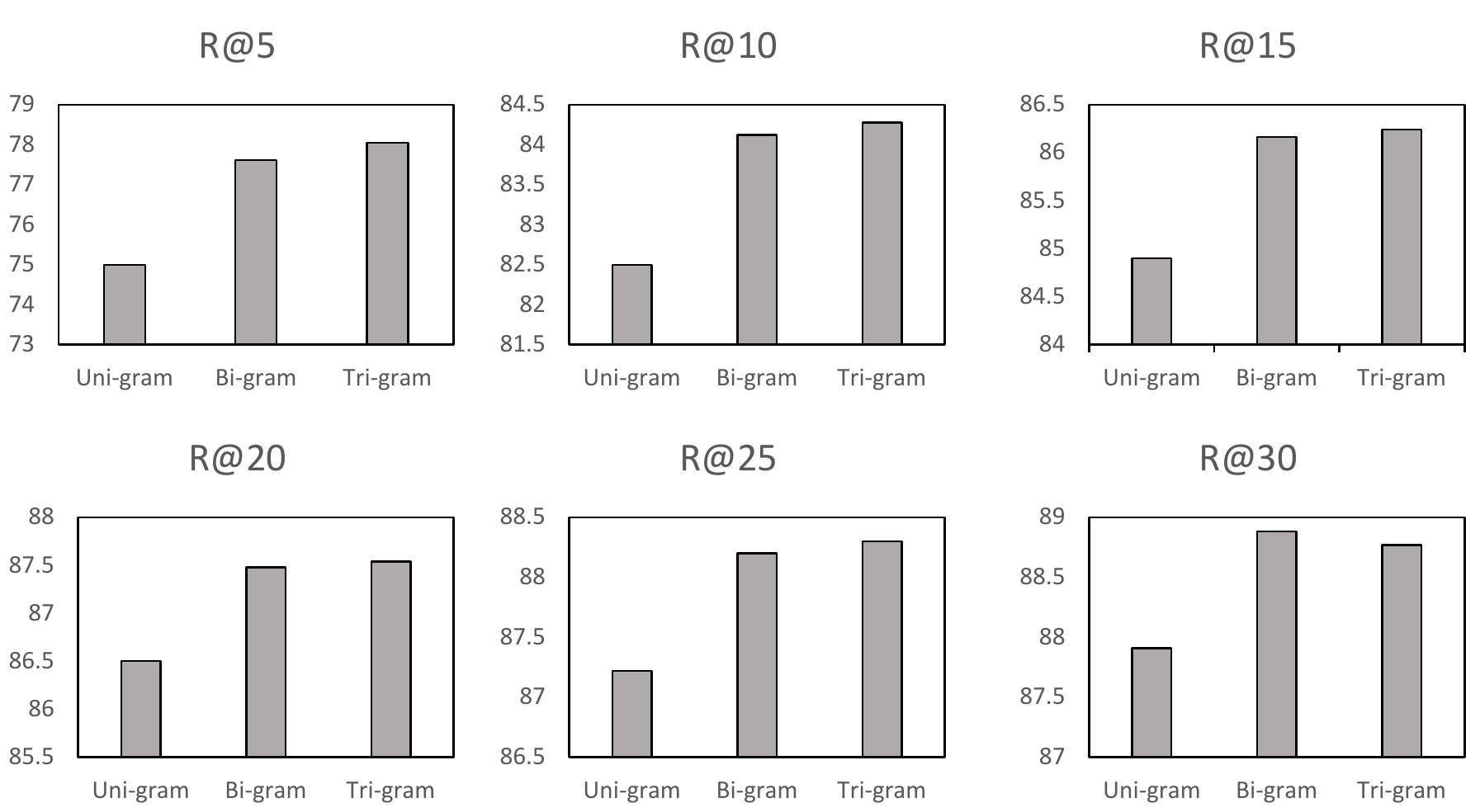}
	\caption{Results of different grams to predict. Improvement is significant by looking ahead one token, but benefit is limited by looking ahead more.}\label{pic.result}
\end{figure}

Firstly, we discuss the choice of $n$ with Figure \ref{pic.result}. Compared with the Uni-gram model, we obverse that looking ahead one future token significantly improves the results and the benefit of looking further is limited. It is due to the short length of target keywords. Most of the problems could be alleviated even with one token to look ahead. We can also see in the case study section\ref{chapter.case} that one length noise token is common for keywords extension. Thus we do not carry experiments for $n \geq 4$. 

Secondly, we discuss options for the residual weight of $\lambda$. We conduct results for a Bi-gram model with $\lambda$ equals 0.4, 0.6, 0.8. Results can be seen from Table \ref{tb:result.lambda}. We observe that using $\lambda=0.6$ or $\lambda=0.8$ reaches comparable results. This result is reasonable. Firstly, $\lambda=0.6$ or 0.8 reaches the balance between maintaining sufficient representation for the decoding token and using future information to assist. Further, no matter what value $\lambda$ is, it is used to modify the ranking score rather than real sentence score, thus as long as one sequence is put into the alive buffer, the same NLG model-consistent sentence score is recorded. Thus our strategy is robust to the choice of hyper-parameter $\lambda$.In other chapters of the paper, explicit n-gram strategies uses $\lambda$ as 0.8.
\begin{table}[h]
	\centering
	\begin{tabular}{|l|c|c|c|c|c|c|}
		\hline
		\textbf{$\lambda$}    & \textbf{R@5} & \textbf{R@10} & \textbf{R@15} & \textbf{R@20} & \textbf{R@25} & \textbf{R@30}\\ \hline
		0.4     & 76.31   & 82.03   & 84.23  & 85.54 & 86.22 & 86.89  \\
		0.6    & 78.13   & 84.09   & 86.07  & 87.23 & 87.89 & 88.65    \\
		0.8   & 77.54   & 84.14   & 86.15  & 87.44 & 88.17 & 88.88    \\ \hline
	\end{tabular}
	\caption{Results for different residual weight $\lambda$ for a Bi-gram model.}\label{tb:result.lambda}
\end{table}

\subsection{Case Study}\label{chapter.case}

In this section, we discuss on how ProphetNet-Ads helps to solve the problems in the generative retrieval model with actual cases. We list three examples that the best baseline model, Tri-gram ProphetNet, failed to find golden ads keywords with the beam size 30 and our model could successfully generate with the beam size 5. 
\begin{table}[h]
\centering
\addtolength{\tabcolsep}{-0.1pt}
\begin{tabular}{|l|l|}
\hline
\multicolumn{1}{|l|}{}                                                                                                       & Baseline:                  \\
\multicolumn{1}{|l|}{\multirow{-2}{*}{\begin{tabular}[c]{@{}l@{}}input:\\ lone wolf discount code\end{tabular}}}             & lone wolf coupon code      \\ \cline{1-1}
\multicolumn{1}{|l|}{}                                                                                                       & lone wolfs                 \\
\multicolumn{1}{|l|}{\multirow{-2}{*}{\begin{tabular}[c]{@{}l@{}}golden:\\ lone wolf distributors coupon code\end{tabular}}} & coupon code discount       \\ \cline{1-1}
Ours:                                                                                                                        & lone wolf car rentals      \\
lone wolf coupon code                                                                                & coupon code coupon code    \\
{\color[HTML]{FE0000} lone wolf distributors coupon code}                                                                    & coupon code contact        \\
lone wolf distributors discount code                                                                                         & ...                        \\
lone wolf distributors promotional code                                                                                      & coupon code pet well being \\
lone wolf distributors promotional codes                                                                                     & coupon code athleta yoga  \\ \hline
\multicolumn{1}{|l|}{}                                                                                         & Baseline:                       \\
\multicolumn{1}{|l|}{\multirow{-2}{*}{\begin{tabular}[c]{@{}l@{}}input:\\ kalathil resort\end{tabular}}}       & kalathil resort                 \\ \cline{1-1}
\multicolumn{1}{|l|}{}                                                                                         & kalamata resort                 \\
\multicolumn{1}{|l|}{\multirow{-2}{*}{\begin{tabular}[c]{@{}l@{}}golden:\\ kalathil lake resort\end{tabular}}} & kalahari hotel                  \\ \cline{1-1}
Ours:                                                                                                          & resort kalahari                 \\
 kalathil resort                                                                        & khao resort                     \\
{\color[HTML]{FE0000} kalathil lake resort}                                                                    & koh samui resorts               \\
kalathi lake resorts                                                                                           & ...                             \\
kalathil lake                                                                                                  & khao lak resort khao lak hotel  \\
kalathil lake resort india                                                                                     & koh samui all inclusive holiday
\\ \hline
\cline{1-1}
\multicolumn{1}{|l|}{\multirow{2}{*}{\begin{tabular}[c]{@{}l@{}}input:\\ workmans car insurance\end{tabular}}}  & Baseline:                      \\
\multicolumn{1}{|l|}{}                                                                                          & \footnotesize{workmans auto insurance quote}  \\ \cline{1-1}
\multicolumn{1}{|l|}{\multirow{2}{*}{\begin{tabular}[c]{@{}l@{}}golden:\\ workmen auto insurance\end{tabular}}} & worx products                  \\
\multicolumn{1}{|l|}{}                                                                                          & walmart car insurance rates    \\ \cline{1-1}
Ours:                                                                                                           & workman islington              \\
{\color[HTML]{FE0000} workmen auto insurance}                                                                                         & walmart auto insurance quote   \\
workmens auto car insurance                                                                                     & walmart auto insurance toronto \\
car insurance man                                                                                               & ...                            \\
car insurance driver women                                                                                      & worxs website call             \\
workmans auto insurance quote                                                                                   & worx warranty registration usa \\ \hline
\end{tabular}
\caption{ Extensions of queries from ProphetNet-Ads and baseline model.
}\label{tb:case.whole}
\end{table}

In the first case of "lone wolf discount code", baseline model fails on generating the desired keyword with the prefix "lone wolf distributors". "distributors" in this case is a noise token for NLG model and baseline model fails to skip the noise. Meanwhile, baseline model search space is filled with the common prefix "coupon code" and finally ending in a range of low-scored outputs because "coupond code" does not have "lone wolf" related suffixes. Baseline model will never achieve "lone wolf discount" with an increasing beam size in this scenario unless we cut Trie's "coupon code" brunch and foresee that "lone wolf distributors" prefix will contain correct information in future tokens. Looking ahead strategies assist in avoiding a optimal local trap in a generative retrieval model, skipping the noise token "distributors", and finally generates all five extensions reasonable.

In the second case of keywords extensions of "kalathil resort", we can see that "kalathil" actually means "kalathil lake" in India. However, "kalathil" is a lake which is an unknown information for a generative retrieval model. Baseline method generates a lot of extensions resembling the input query, but most of them are wrong. Our model implicitly knows the combination of "kalathil lake", by looking ahead. Looking ahead strategies allow generative retrieval model to find a proper path with golden target information to go.

In the last case of keywords extensions of "workmans car insurance", two difficulties are there for generative retrieval models: "workmen" is misspelled as "workmans" and synonym words "car" and "auto" used in the query and outputs. Both of the models are powerful enough to learn that "car" and "auto" are synonymous, but baseline model fails in generating "workmen". It is because no sufficient data about misspelled "workmans" and correct "workmen" are provided in the training corpus. But our model successfully generate it by looking ahead future information. Other extensions are also more reasonable than baseline ones, with diverse prefix "workmen", "workmans" and "car insurance", which also show the strong retrieval ability of our model.

\section{Conclusion}
In this work, we investigate the weakness of present generative retrieval models and propose ProphetNet-Ads to improve the retrieval ability. For the experiments, we collect a keywords extension dataset from a real-world search engine. We carry experiments on the recently proposed Trie-based LSTM generation model and other variants of generative retrieval models to analyze generative retrieval models in keywords extension task. Experimental results show that ProphetNet-Ads brings significant improvement over the recall and MAP metrics. 

\bibliographystyle{splncs04}
\bibliography{ngram_ref}

\end{document}